\documentclass[10pt,conference]{IEEEtran}
\usepackage{cite}
\usepackage{amsmath,amssymb,amsfonts}
\usepackage{algorithm}
\usepackage{algorithmic}
\usepackage{graphicx}
\usepackage{booktabs}
\usepackage{textcomp}
\usepackage{xcolor}
\usepackage[hyphens]{url}
\usepackage{fancyhdr}
\usepackage{hyperref}
\usepackage{xspace}
\usepackage{CJKutf8}

\newcommand{\hpcayear}{2027}

\newcommand{\sysname}{WaveAlign\xspace}
\newcommand{\para}[1]{\noindent\textbf{#1}}

\title{\sysname: Cache-Aware Query-Row Scheduling for Sparse Attention in Long-Video Generation}

\newcommand\hpcaauthors{
\begin{tabular}{c}
Zijian Dai$^{1,2}$,
Sen Han$^{1}$,
Youhui Bai$^{1}$,
Shannon Wang$^{2}$,
Kan Wu$^{1}$ \\[1pt]
Jingkai Huang$^{3}$,
Yuhang Wang$^{1}$,
Jing Li$^{1}$,
Cheng Li$^{1,2}$
\end{tabular}
}

\newcommand\hpcaaffiliation{
$^{1}$University of Science and Technology of China \\
$^{2}$Institute of Artificial Intelligence, Hefei Comprehensive National Science Center \\
$^{3}$South China University of Technology
}

\newcommand\hpcaemail{
daizijian2080@outlook.com,
handsome2022@mail.ustc.edu.cn,
youhuibai@ustc.edu.cn,
wsn511799@163.com,\\
kanwu.wisc@gmail.com,
202364820481@mail.scut.edu.cn,
wyh2022@mail.ustc.edu.cn,
lj@ustc.edu.cn,
chengli7@ustc.edu.cn
}

\author{
    \IEEEauthorblockN{\hpcaauthors{}}
    \IEEEauthorblockA{
        \hpcaaffiliation{} \\
        \hpcaemail{}
    }
}

\fancypagestyle{camerareadyfirstpage}{%
  \fancyhead{}
  
  \fancyhead[C]{
    \ifdefined\aeopen
    \parbox[][12mm][t]{13.5cm}{\hpcayear{} IEEE International Symposium on High-Performance Computer Architecture (HPCA)}    
    \else
      \ifdefined\aereviewed
      \parbox[][12mm][t]{13.5cm}{\hpcayear{} IEEE International Symposium on High-Performance Computer Architecture (HPCA)}
      \else
      \ifdefined\aereproduced
      \parbox[][12mm][t]{13.5cm}{\hpcayear{} IEEE International Symposium on High-Performance Computer Architecture (HPCA)}
      \else
      \parbox[][0mm][t]{13.5cm}{\hpcayear{} IEEE International Symposium on High-Performance Computer Architecture (HPCA)}
    \fi 
    \fi 
    \fi 
    \ifdefined\aeopen 
      \includegraphics[width=12mm,height=12mm]{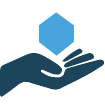}
    \fi 
    \ifdefined\aereviewed
      \includegraphics[width=12mm,height=12mm]{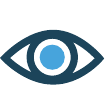}
    \fi 
    \ifdefined\aereproduced
      \includegraphics[width=12mm,height=12mm]{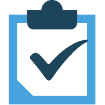}
    \fi
  }
  \fancyfoot[C]{}
}
\begin{document}
\maketitle

\ifdefined\hpcacameraready 
  \thispagestyle{camerareadyfirstpage}
  \pagestyle{empty}
\else
  \thispagestyle{plain}
  \pagestyle{plain}
\fi

\newcommand{\hpcaheight}{0mm}
\ifdefined\eaopen
\renewcommand{\hpcaheight}{12mm}
\fi

\begin{abstract}
Long-video generation with diffusion transformers (DiTs) produces extremely long token sequences, making attention a dominant inference bottleneck. Dynamic sparse attention reduces computation, but its realized speedup remains limited because irregular query-row execution degrades L2 cache locality and increases HBM traffic. 

We present \sysname, a lightweight, cache-aware query-row reordering framework for dynamic sparse attention. 
\sysname formulates row ordering as an optimization problem and approximates it with two stages. 
The first stage derives a low-rank SVD representation of sparse-mask rows and groups query rows with similar K/V access patterns, increasing K/V overlap among concurrently scheduled rows. The second stage exploits streaming GPU scheduling by sorting rows within each wave in descending order of their K/V-block counts, so that short rows from the current wave are followed by long rows from the next. This aligns K/V accesses across wave boundaries and enables shared blocks to be reused before eviction.
An adaptive skip module avoids unprofitable reordering. By only permuting query and mask rows, \sysname preserves sparse-attention semantics and requires no changes to existing methods or backend kernels.
Across two GPU architectures, two video DiTs, and four sparse-attention methods, \sysname raises the L2 cache hit ratio from 28.48\%--36.35\% to 79.38\%--89.06\%, reduces HBM read traffic by up to 92.11\%, and achieves up to $1.25\times$ kernel and $1.17\times$ end-to-end generation speedup without quality loss.

\end{abstract}

\section{Introduction}
Diffusion transformers (DiT) has emerged as a dominant paradigm for high-fidelity video generation, for its strong scalability and high-fidelity synthesis capabilities~\cite{dit}. Compared with image generation, video generation typically requires handling longer sequences, since the sequence length increases with both spatial resolution and the number of frames, so video generation remains very costly and time-consuming. The bottleneck mainly comes from the quadratic complexity in terms of input sequence length in the attention module~\cite{flashattention}.  For example, generating a 401-frame 720p video, corresponding to a 364K sequence length, takes 10,678 seconds using Wan2.1-14B~\cite{wan21_t2v_14b_modelcard} on a single NVIDIA H100 GPU, among which the attention accounts for 85.86\%. Similar behaviors are also observed in LTX2.3~\cite{ltx23} and other video DiTs.

Utilizing the inherent sparsity of attention~\cite{minference,flexprefill,xattention,spargeattention}, each query (block) only selects its most relevant KV (blocks)  for attention calculation, which  significantly reduces the attention computation while preserving generation quality. Nevertheless, in practice, the speedup brought by such sparse attention methods is far below the reduction in computation, even with GPU-optimized backend kernels such as FlashAttention4~\cite{flashattention4} and FlashInfer~\cite{flashinfer}. Across Wan2.1 and LTX2.3, the speedup can only reach 80$\%$ of the expected level. This gap reveals substantial untapped optimization potential in existing GPU implementations of sparse attention.

Our profiling shows that this gap is fundamentally a memory-locality
problem. Sparsification increases the L2 miss rate, collapses L2 cache reuse, and drives HBM
bandwidth toward saturation, shifting attention from compute-bound to
memory-bandwidth-bound execution. We trace this transition to two root causes of poor L2 cache locality. A \textit{wave} comprises Q blocks executing concurrently across all GPU compute units, together with the K/V blocks they access. First, Q blocks in the same
wave select spatially scattered, weakly overlapping K/V blocks, so their
union overflows L2 cache and blocks are evicted before reuse. 
Second, Q blocks select different numbers of K/V blocks, creating workload imbalance that causes adjacent waves to interleave in execution, temporally misaligning accesses to shared blocks and extending their reuse distance.

To address these problems, we present \sysname, a lightweight high-performance sparse attention framework. Note that the attention outputs w.r.t. the queries are computed independently. Through reordering the query rows, gathering the rows accessing similar KV blocks, \sysname improves L2 locality and hence improves the kernel efficiency. 
However, finding an effective row order is non-trivial.
Existing locality-aware reordering methods for sparse matrix multiplication~\cite{jiang_lsh_spmm,hypergraph_spgemm,chierichetti_shingle,bootes} cannot be directly applied to online sparse attention. 
Lower-cost methods provide little improvement in L2 locality, whereas methods that substantially improve locality incur preprocessing overhead that outweighs their kernel-time savings. 
A practical solution must capture similarities across irregular K/V access patterns and mitigate row-workload imbalance while providing reordering overhead from offsetting the resulting kernel gains.

To this end, \sysname proposes a cache-aware query-row reordering method, which cast row reordering as an optimization problem and
approximate its solution with a lightweight two-stage algorithm.
Exploiting the observation that similarities among irregular mask rows
are low-rank, Stage~1 embeds the rows via a matrix-free rank-2
SVD~\cite{golub2013matrix} and orders them by polar angle, grouping rows with similar
K/V-access patterns. Stage~2 then sorts rows within each wave by
descending K/V-block count, so that short and long rows interleave
across wave boundaries and shared K/V blocks are reused before eviction.
The two stages are complementary: Stage~1 creates cross-row reuse
opportunities and Stage~2 realizes them before cache eviction, jointly
improving L2 locality. WaveAlign only permutes
query and mask rows and restores the original output order,
preserving sparse-attention semantics without modifying the
sparse algorithms or backend kernels. Moreover, \sysname introduces an automatic discrimination mechanism
that bypasses query-row reordering when the expected gain cannot amortize its overhead, such as for short sequences or highly sparse masks.

We implement \sysname in vLLM-Omni~\cite{vllm_omni} and evaluate it on two GPU architectures, using FlashAttention4 as backend kernel on NVIDIA H100 and FlashInfer on NVIDIA A40. 
Our evaluation spans two widely used video DiT families, Wan2.1~\cite{wan21_repo} and LTX2.3~\cite{ltx23}, and four representative dynamic sparse-attention methods. 
Across these workloads, \sysname increases the L2 cache hit ratio from 28.48\%--36.35\% to 79.38\%--89.06\% and reduces HBM read traffic by 80.71\%--92.11\%, shifting sparse attention from the memory-bandwidth-bound regime back into the compute-bound regime. This improved memory efficiency yields up to $1.25\times$ kernel speedup, which translates into up to $1.17\times$ end-to-end generation speedup. Because \sysname preserves sparse-attention semantics, it maintains the same generation quality as the corresponding baselines.
\section{Background}
\subsection{Video Diffusion Transformers (DiTs)}
\label{subsec:video-dit}

\para{DiT models.} 
Video generation powers a growing range of applications, from entertainment and advertising to virtual reality~\cite{videosurvey,xing2025survey}.
Video diffusion transformers (DiTs)~\cite{latte,cogvideox,wan21_repo}, which pair diffusion models with Transformer backbones~\cite{dit}, have become the dominant architecture, as they scale well and excel at capturing long-range spatiotemporal dependencies~\cite{cogvideox,wan21_repo}.

Popular video DiTs share the same recipe. 
A video is encoded into a sequence of spatiotemporal tokens, whose length grows with both the frame count and the spatial resolution.
Each Transformer layer applies multi-head attention (MHA) to capture dependencies across tokens, followed by a feed-forward network (FFN)~\cite{transformer}. 
MHA projects the hidden states $X$ into queries ($Q$), keys ($K$), and values ($V$), computes softmax-normalized query--key scores, and aggregates the values accordingly~\cite{transformer}.
Unlike autoregressive large language models (LLMs), which restrict attention to preceding tokens with a causal mask, video DiTs apply full, non-causal self-attention over the entire sequence. Inference iteratively refines the latent noise over tens of denoising steps, each requiring a complete forward pass~\cite{ddpm}. Because hidden states change at every step, the derived K/V tensors are transient and cannot be cached across steps, forcing full long-sequence attention to be recomputed throughout inference.

\para{Computation bottleneck of DiT.}
Despite their generation quality, DiTs are painfully slow.
Using Wan2.1 T2V 1.3B~\cite{wan21_t2v_1p3b_modelcard}, a popular open source DiT model, generating an 81 frame video at a resolution of 720$\times$1280 takes over five minutes on an H100 GPU. At 1920$\times$1080, the generation time increases dramatically to approximately 25 minutes~\cite{dcvideogen}.

Our profiling shows that attention dominates the generation time, accounting for 73\% and 90\% in the two settings above. %
This overhead stems from the quadratic complexity of attention with respect to sequence length~\cite{flashattention}. Video sequences are exceptionally long and grow with both frame count and spatial resolution, reaching 76K and 364K tokens in these settings. 
Moreover, bi-directional attention and iterative denoising incur this full quadratic cost at every layer and denoising step. 
Other popular video DiTs, such as LTX2.3~\cite{ltx23}, exhibit similar behavior. Therefore, reducing attention cost is essential for  efficient long, high-resolution video generation.

\subsection{Sparse Attention for Long-Video Generation}
\label{subsec:video-attention-background}

To mitigate the high cost of full attention, recent studies exploit the inherent sparsity of attention mechanisms~\cite{svg}.
Only a small subset of key/value tokens typically receives substantial attention weights and materially affects the output.
Identifying these tokens and computing only the corresponding interactions can substantially reduce computation while preserving generation quality~\cite{sun2026efficient}.

Consequently, recent works have increasingly favored \textit{dynamic-pattern} methods, as important interactions vary across inputs. These methods estimate relevance from the online attention statistics and select important K/V tokens for each query at runtime, offering broader applicability and better accuracy preservation. 
MInference~\cite{minference} instantiates head-specific A-shape, vertical-slash, or general sparse patterns; FlexPrefill~\cite{flexprefill} further adapts the pattern and budget to each input and head; XAttention~\cite{xattention} scores tokens via antidiagonal aggregation; and SpargeAttn~\cite{spargeattention} predicts relevance from pooled Q/K features and refines it via online softmax statistics.

\begin{figure}[t!]
    \centering
    \includegraphics[width=1\linewidth]{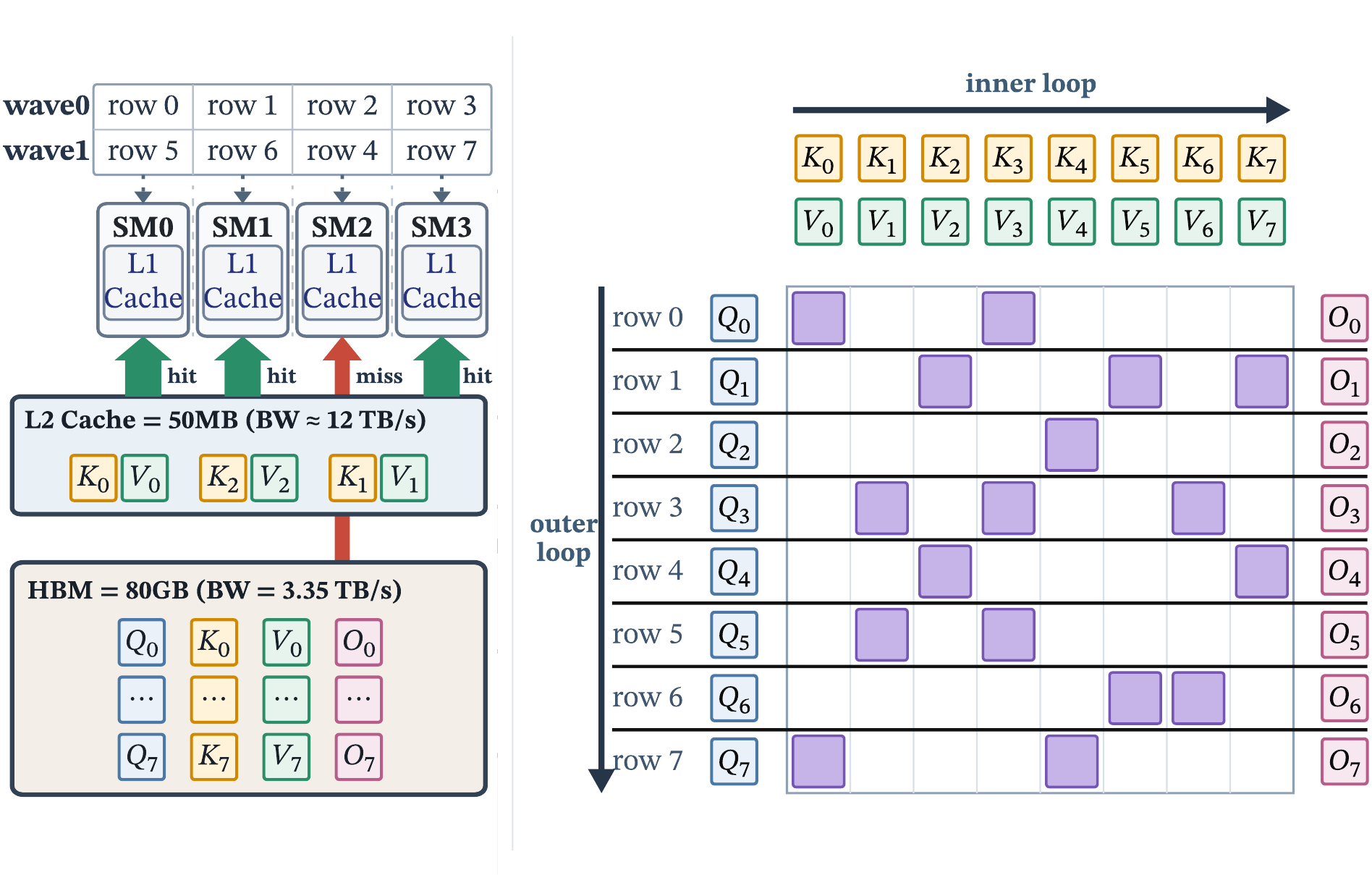}
    \caption{GPU architecture, block-sparse attention, and the mapping of query rows and K/V blocks onto GPU execution units. We use NVIDIA H100 as an example and show 4 of its 132 SMs for clarity.}
    \label{fig:sm-schedule}
\end{figure}

\para{Block-sparse attention for modern GPUs.}
To exploit the GPU memory hierarchy as shown on the left of Figure~\ref{fig:sm-schedule}, memory accesses should be contiguous, while on-chip computation is naturally tiled over data blocks~\cite{flashattention,flashattention2}. GPU-oriented dynamic sparse attention therefore organizes the $Q$, $K$, and $V$ tensors and their computation at block granularity, with each block containing tens to hundreds of consecutive tokens.

Based on this organization, dynamic sparse attention encodes block-level computation using a binary mask $M$. Dense attention computes each query block $Q_i$ with all K/V block pairs to produce $O_i$, whereas sparse attention executes only the interactions retained by $M$. 
As illustrated on the right of Figure~\ref{fig:sm-schedule}, each mask row corresponds to a query block. A purple block with $M_{ij}=1$ denotes an executed interaction between $Q_i$ and $(K_j,V_j)$, while a white block with $M_{ij}=0$ denotes a skipped interaction. For example, $Q_0$ uses $(K_0,V_0)$ and $(K_3,V_3)$ to produce $O_0$, while $Q_3$ also uses $(K_3,V_3)$, showing that different query blocks may share K/V blocks. The row sum $n_i=\sum_j M_{ij}$, or its number of nonzero blocks (NNZ), determines the workload of $Q_i$.

\subsection{Mapping Sparse Attention onto GPUs}
\label{subsec:gpu-sparse-execution}

After constructing the block mask $M$, the sparse-attention method passes $Q$, $K$, $V$, and $M$ to a backend kernel such as FlashAttention4~\cite{flashattention4} or FlashInfer~\cite{flashinfer} for attention computation. 
As illustrated in Figure~\ref{fig:sm-schedule}, the $QKV$ tensors and attention output $O$ reside in HBM, while the kernel loads the required blocks on chip for computation. 
These kernels follow the FlashAttention-style nested-loop structure, with query blocks in the outer loop and the K/V blocks selected by the corresponding mask row in the inner loop. 
For each $Q_i$, the task represented by mask row $i$ is assigned to one streaming multiprocessor (SM) on the GPU, which maintains the online-softmax state and partial output $O_i$ on chip while iterating over its selected K/V blocks. The completed $O_i$ is then written back to HBM.

The query-row tasks concurrently occupying all SMs form a \texttt{wave}~\cite{nvidia_nsight_compute_profiling_guide}. In the four-SM example in Figure~\ref{fig:sm-schedule}, rows 0--3 initially form wave 0, followed by rows 4--7 in wave 1. 
A wave is not a synchronization barrier. If row 2 finishes first because it accesses fewer K/V blocks, the freed SM immediately begins row 4 while the remaining rows of wave 0 continue. 
Because all SMs share the L2 cache, a K/V block loaded by one row can be reused by another if it remains resident in L2 cache; otherwise, it must be fetched again from HBM. 
The K/V access overlap, row workloads, and execution order therefore jointly determine the L2 cache hit ratio and HBM traffic.

\section{Motivation}
\label{sec:motivation}

Sparse attention eliminates a substantial fraction of the FLOPs in dense attention, yet existing implementations often fail to achieve proportional speedups. We first quantify this gap, then show that sparsification shifts the kernel from compute-bound to memory-bandwidth-bound (Section~\ref{subsec:mem-bandwidth-bound}), and finally identify two root causes (Section~\ref{subsec:two-root-cause}).

\begin{figure}[t]
    \centering
    \includegraphics[width=1\linewidth]{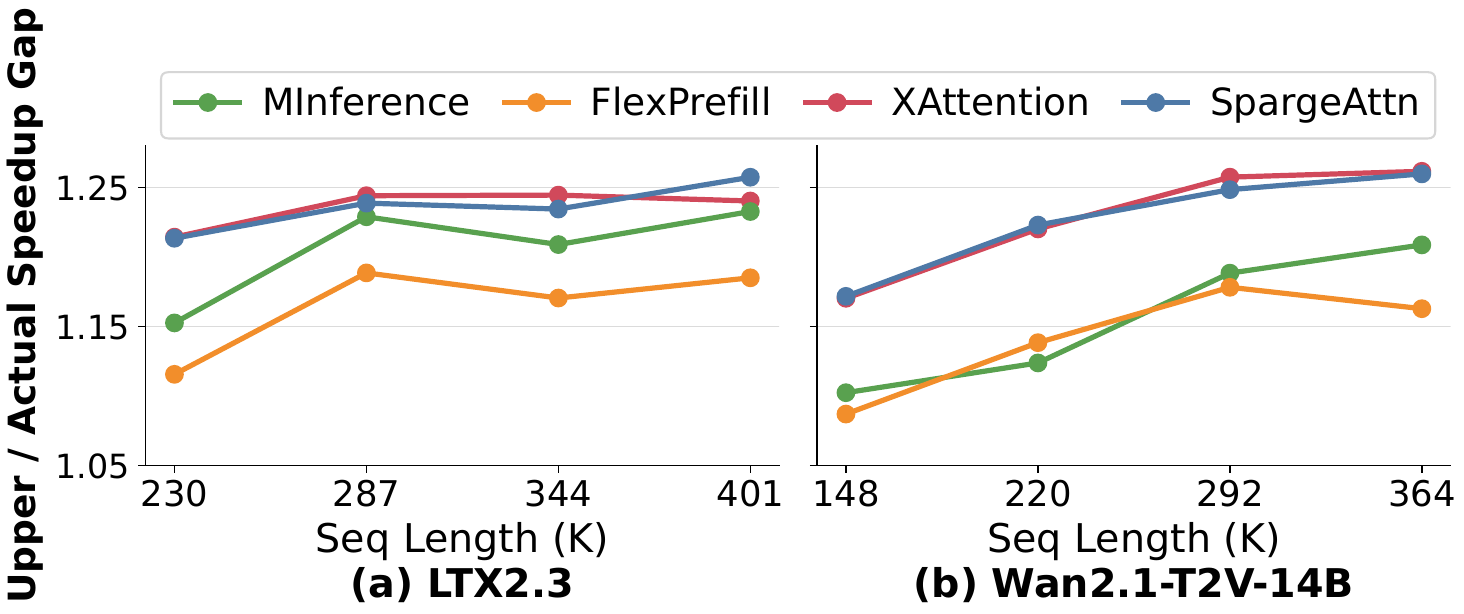}
    \caption{Upper bound vs. actual attention kernel speedup across sequence lengths on Wan2.1-T2V-14B and LTX2.3, with each sparse-attention method using its default sparsity configuration on H100.}
    \label{fig:ltx_wan_gap}
\end{figure}

\subsection{FLOP Reduction Does Not Yield Proportional Speedup}

We evaluate four representative dynamic sparse-attention methods: MInference~\cite{minference}, FlexPrefill~\cite{flexprefill}, XAttention~\cite{xattention}, and SpargeAttn~\cite{spargeattention}.
We run them on two widely used video DiT models, Wan2.1-T2V-14B~\cite{wan21_t2v_14b_modelcard} and LTX2.3~\cite{ltxvideo}, using an NVIDIA H100 GPU with FlashAttention4~\cite{flashattention4} as the backend kernel.
The workloads span sequence lengths of 148K--364K tokens for Wan2.1 and 230K--401K tokens for LTX2.3; detailed setups are provided in Section~\ref{subsec:evaluation-setup}. 
For each workload, $\rho$ denotes the retained block density, i.e., the fraction of non-skipped blocks among all dense-attention blocks. Executing a $\rho$ fraction of the blocks gives a compute-proportional upper-bound speedup of $1/\rho$. We measure the actual sparse-kernel speedup over the dense kernel as $S_{\mathrm{actual}}$, and define the gap as $(1/\rho)/S_{\mathrm{actual}}$.

Figure~\ref{fig:ltx_wan_gap} shows a consistent gap between the compute-proportional and measured speedups. The upper bound exceeds the measured speedup by 1.12--1.26$\times$ on LTX2.3 and 1.09--1.26$\times$ on Wan2.1, with the gap generally increasing at longer sequence lengths. 
MInference and FlexPrefill, whose masks retain predefined structures, exhibit smaller gaps than XAttention and SpargeAttn, which construct fully content-dependent dynamic masks. These results show that FLOP reduction alone cannot yield sparse-attention kernel speedup.

\begin{table}[t]
  \centering
  \caption{NCU characterization on attention module of Wan2.1-14B model with 364K tokens on H100. }
  \label{tab:motivation-memory-bound2}

  \normalsize
  \setlength{\tabcolsep}{3.0pt}
  \renewcommand{\arraystretch}{0.8}

  \begin{tabular*}{\columnwidth}{
    @{\extracolsep{\fill}}lccc@{}
  }
    \toprule
    Methods
    &
    \begin{tabular}[c]{@{}c@{}}
      Density $\rho$ $(\%)$
    \end{tabular}
    &
    \begin{tabular}[c]{@{}c@{}}
      HBM SOL $(\%)$
    \end{tabular}
    &
    \begin{tabular}[c]{@{}c@{}}
      L2 Hit $(\%)$
    \end{tabular}
    \\
    \midrule
Dense
      & $100.0$
      & $49.38$
      & $46.51$ \\
    MInference
      & $18.3$
      & $79.28$
      & $21.67$ \\
    FlexPrefill
      & $29.4$
      & $86.43$
      & $14.72$ \\
    XAttention
      & $28.8$
      & $83.44$
      & $17.94$ \\
    SpargeAttn
      & $28.3$
      & $84.06$
      & $17.35$ \\
    Random
      & $20.0$
      & $87.77$
      & $10.64$ \\
    \bottomrule
  \end{tabular*}
\end{table}

\subsection{Sparse Attention Becomes Memory-Bandwidth-Bound}
\label{subsec:mem-bandwidth-bound}

To locate the source of the speedup gap, we profile the 364K-token Wan2.1 workload from Figure~\ref{fig:ltx_wan_gap} using NVIDIA Nsight Compute (NCU)~\cite{nsight_compute_2025_3_1}.
Table~\ref{tab:motivation-memory-bound2} reports each method's default density $\rho$, HBM bandwidth saturation measured as Speed of Light (SOL)~\cite{nvidia_nsight_compute_profiling_guide}, and L2 hit ratio.
The results show that sparsification degrades cache behavior.
Dense attention achieves the highest L2 hit ratio and the lowest HBM SOL, because its regular K/V access stream enables effective cache reuse.
All four dynamic sparse methods, in contrast, issue irregular and discontinuous K/V accesses, which substantially reduce L2 locality and drive up HBM SOL.
Consistent with their smaller speedup gaps in Figure~\ref{fig:ltx_wan_gap}, MInference and FlexPrefill retain more structured access patterns and therefore better L2 locality than XAttention and SpargeAttn.
Finally, a synthetic random mask approximates the locality floor for fully unstructured K/V accesses, which XAttention and SpargeAttn already approach.

\begin{figure}[t]
    \centering
    \includegraphics[width=1\linewidth]{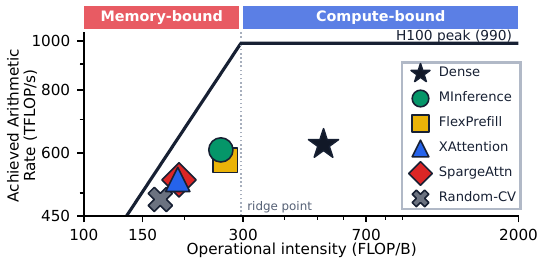}
    \caption{Roofline analysis on attention kernels of Wan2.1-14B model with 364K tokens on H100.}
    \label{fig:roofline471K}
\end{figure}

To determine whether the increased HBM traffic shifts the kernel bottleneck from computation to memory bandwidth, we further conduct a roofline analysis~\cite{williams2009roofline} in Figure~\ref{fig:roofline471K}. 
We estimate FLOPs from the two matrix multiplications in attention using the number of executed attention blocks and the per-block operation count. Kernel execution time and HBM traffic are measured directly with NCU tool. 
The achieved arithmetic throughput is calculated by dividing the estimated FLOPs by the measured kernel time, while operational intensity is calculated by dividing the estimated FLOPs by the measured HBM traffic. The compute roof corresponds to the theoretical FP16 peak throughput of H100. 
The results show that dense attention is compute-bound, whereas all sparse methods become memory-bandwidth-bound. Although sparsification reduces FLOPs, poor L2 locality prevents HBM traffic from decreasing proportionally, limiting kernel speedup and producing the gap observed in Figure~\ref{fig:ltx_wan_gap}.

\begin{figure}
    \centering
    \includegraphics[width=1\linewidth]{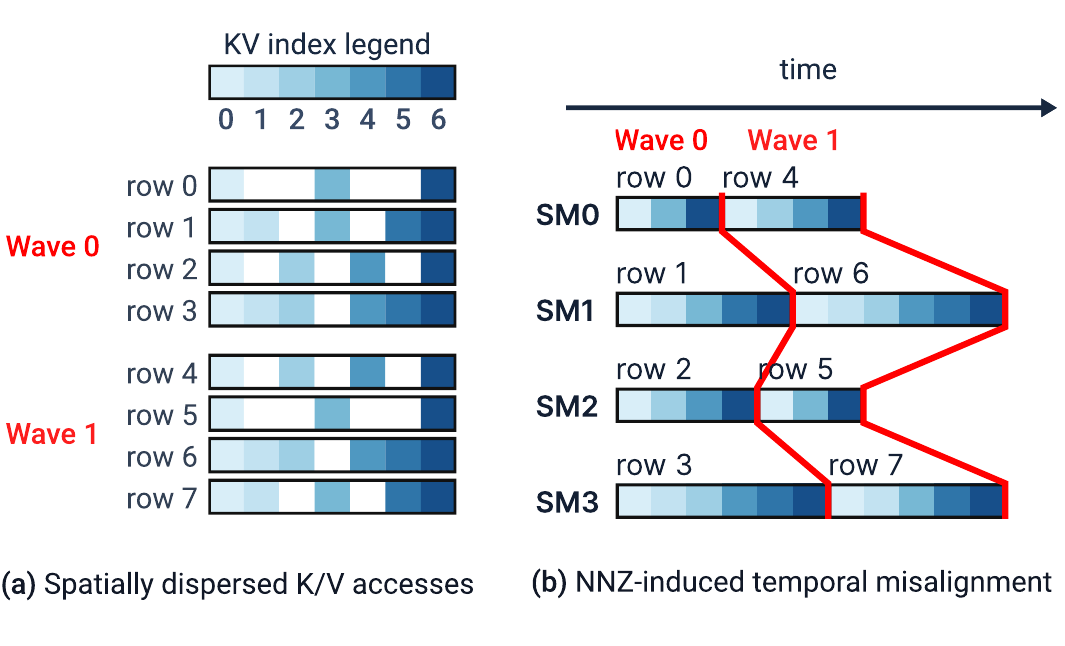}
    \caption{Two root causes of poor L2 cache locality.}
    \label{fig:spatial_failure}
\end{figure}

\subsection{Why Access Is Irregular: Two Root Causes}
\label{subsec:two-root-cause}

Effective L2 reuse requires rows executing within the same wave to access overlapping K/V blocks and to reach these blocks close together in time. Dynamic masks undermine both conditions. As illustrated in Figure~\ref{fig:spatial_failure}, we consider eight query rows scheduled on four SMs, where rows 0--3 form Wave 0 and rows 4--7 form Wave 1. Each colored block denotes an accessed K/V block, with its color indicating the block index.

\para{Root cause 1: Spatially dispersed K/V accesses.}
Rows scheduled in the same wave share few K/V blocks, because dynamic masks assign each row an irregular access set (Figure~\ref{fig:spatial_failure}(a)).
With little overlap, the SMs' concurrent accesses span a large union of K/V blocks that overflows the working set of L2 cache.
Useful blocks are evicted before they are reused, and each eviction converts a potential L2 hit into an extra HBM fetch.

\para{Root cause 2: NNZ-induced temporal misalignment.}
Even rows with overlapping access sets may fail to reuse shared blocks when their accesses are separated in time. As shown in Figure~\ref{fig:spatial_failure}(b), rows access different numbers of K/V blocks and therefore take different times to execute. Under streaming scheduling, an SM that finishes a shorter row immediately starts one from the next wave, temporally interleaving the two waves. This workload imbalance separates accesses to shared blocks. Although rows 4--7 all access the block pair $(K_6,V_6)$, they reach it far enough apart that the cached copy may be evicted between accesses, forcing repeated HBM fetches.

These two root causes jointly determine whether K/V reuse can occur. The sparse method fixes which K/V blocks each row needs, but not which rows run together or when they reach the shared blocks.
Spatial overlap creates opportunities for cross-row reuse, while temporal alignment determines whether these opportunities can be realized before the blocks are evicted from L2. 
A failure in either condition leads to poor L2 cache locality and repeated HBM accesses.

\section{\sysname Internals}
\label{sec:design}

\subsection{Design Rationale}

\begin{figure}
    \centering
    \includegraphics[width=1\linewidth]{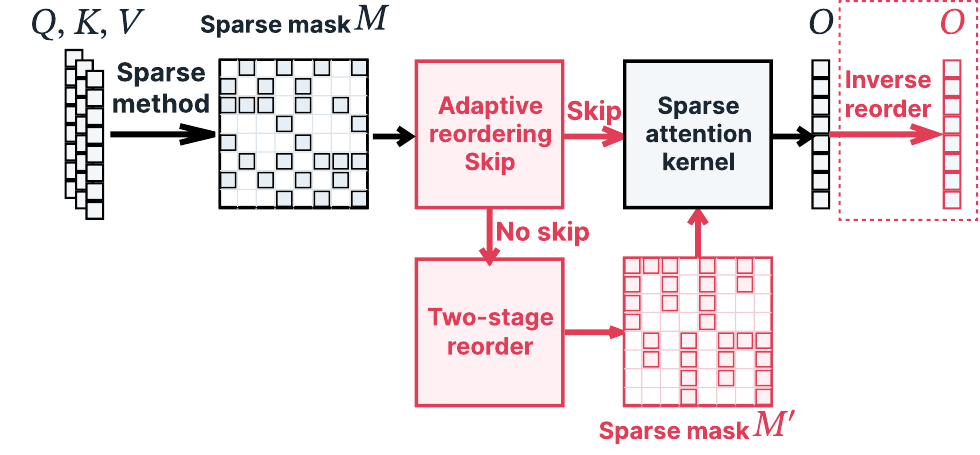}
    \caption{Workflow of WaveAlign.}
    \label{fig:WaveAlign-overview}
\end{figure}

As described in Section~\ref{subsec:gpu-sparse-execution}, each query row is mapped to an SM, while rows in the same wave execute concurrently or in close temporal proximity and share the L2 cache. Meanwhile, query rows are mutually independent, allowing query and mask rows to be jointly reordered and their outputs restored afterward without changing attention semantics. Together, these observations enable us to group rows with overlapping K/V accesses into the same wave, increasing cross-SM L2 reuse and reducing HBM traffic.

Building on this opportunity, we propose \sysname, a lightweight query-row scheduling framework.
As shown in Figure~\ref{fig:WaveAlign-overview}, the original workflow of sparse attention in the black path directly passes mask $M$ to the backend kernel.
In contrast, \sysname introduces the red components as plug-in stages between mask construction and the backend kernel.
Once receiving $M$, \sysname first uses an adaptive guard to determine whether reordering is profitable (Section~\ref{subsec:runtime-coordination}).
If reordering is skipped, the workload follows the original path.
Otherwise, \sysname applies its two-stage row-ordering strategy to construct a reordered mask $M'$ (Section~\ref{subsec:svd-reordering} and ~\ref{subsec:nnz-reordering}).
After kernel execution, an inverse reorder restores the output rows to their original positions.
Since each query row still attends to exactly the same K/V blocks, \sysname changes only the physical execution order and preserves both the semantics of sparse attention.

However, realizing this workflow efficiently poses two challenges. First, the irregular masks generated by dynamic sparse attention make it difficult to efficiently identify rows with similar K/V access patterns. Second, reordering must incur low overhead, since it is repeatedly performed across attention heads, layers, and denoising steps and can otherwise offset the resulting kernel speedup.

\subsection{Problem Definition}
\label{subsec:rowalign-objective}

To systematically address these challenges, we first formulate this problem.
For attention head $h$, let $M_h \in \{0,1\}^{M_B \times N_B}$ denote its block-sparse mask, where $M_B$ and $N_B$ are the numbers of query and K/V blocks, respectively, and $M_h[i,j]=1$ indicates that query block $i$ accesses K/V block $j$. \sysname aims to maximize the K/V-block overlap among query rows assigned to the same scheduling wave.  Specifically, we aim to divide the whole $M_h$ into $k$ waves, and each wave contains at most $W$ query-block rows selected from $M_h$. In order to make the rows within the same wave as close as possible,  we construct the following problem:

\begin{equation} \label{eq:obj}
\begin{aligned}\
    \min_{P_h}\quad
    & \sum_{l=1}^{k}
      \sum_{i,j}
      \left\|S^l_{h}[i] - S^l_{h}[j]\right\|^2 \\
    \mathrm{s.t.}\quad
    & S_h^l = P_h^l M_h, \\
    & P_h = [P_h^1;P_h^2;\ldots;P_h^k], \\
    & P_h \in \{0,1\}^{M_B\times M_B}, \\
    & P_hP_h^{T}=I .
\end{aligned}
\end{equation}

Here, $P_h$ is the row-permutation matrix for head $h$, and $P_h^l$ selects the rows assigned to wave $l$.
Accordingly, $S_h^l$ denotes the reordered mask rows executed in that wave.
The constraint $P_hP_h^T=I$ ensures that each query row appears exactly once.

\subsection{SVD-Based Global Spatial Ordering}
\label{subsec:svd-reordering}

\begin{figure}
    \centering
    \includegraphics[width=1\linewidth]{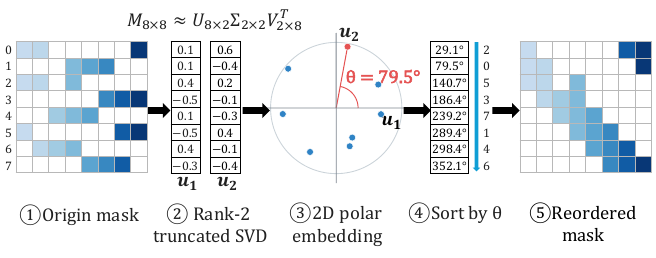}
    \caption{Example of SVD-Based Global Spatial Ordering.}
    \label{fig:Stage1}
\end{figure}

For each attention head, directly searching the optimal permutation requires jointly partitioning and ordering all mask rows to minimize their intra-wave distances, which is NP-hard, motivating the efficient approximation introduced next.

Problem~(\ref{eq:obj})  seeks a row partition that
keeps mutually close rows in the same wave. This is exactly what SVD
exposes.  Since statistically several query blocks attend to nearly identical K/V
blocks, the rows are highly correlated and the singular spectrum is
top-heavy, and a few leading directions already capture the dominant row
geometry. In the following, we show that 
the row partitioning of Problem~(\ref{eq:obj}) can be turned into a
low-dimensional geometric problem in a principled rather than heuristic
way, as illustrated in  Figure~\ref{fig:Stage1}.

Before using the SVD, we first apply column centering to the mask,
\begin{equation}
  \widetilde{M}_h=\Bigl(I-\tfrac{1}{M_B}\mathbf{1}\mathbf{1}^{\top}\Bigr)M_h .
\end{equation}
Centering subtracts the same mean row from every row, so all pairwise
row distances are preserved, namely, 
$\lVert\widetilde{M}_h[i]-\widetilde{M}_h[j]\rVert=\lVert M_h[i]-M_h[j]\rVert$,
so reordering $M_h$ is equivalent to reordering $\widetilde{M}_h$.

Write the SVD as $\widetilde{M}_h=\sum_{t}\sigma_t u_t v_t^{\top}$ with
$\sigma_1\ge\sigma_2\ge\cdots\ge\sigma_R\ge0$ and $R=\min(M_B,N_B)$.
Since the right singular vectors $\{v_t\}$ are orthonormal, the pairwise
row distance decomposes exactly as
\begin{equation}
  \lVert\widetilde{M}_h[i]-\widetilde{M}_h[j]\rVert^{2}
  =\sum_{t}\sigma_t^{2}\bigl(u_t[i]-u_t[j]\bigr)^{2}.
\end{equation}
Assigning row $i$ the coordinate vector
$\phi_r(i)=(\sigma_1 u_1[i],\dots,\sigma_r u_r[i])$ therefore makes the
Euclidean distance in the $r$-dimensional embedding equal to the
rank-$r$ partial sum of the true row distance, with residual
$\epsilon_{ij}=\sum_{t>r}\sigma_t^{2}(u_t[i]-u_t[j])^{2}$, which is
small when the singular spectrum decays quickly, as is typical for
these masks. By the Eckart--Young theorem~\cite{eckart1936approximation,mirsky1960symmetric},
this truncation is the best rank-$r$ approximation of
$\widetilde{M}_h$, so the leading (scaled) left singular vectors are
exactly the $r$-dimensional coordinates that best preserve the pairwise
row distances in the least-squares sense (classical multidimensional
scaling). The SVD thus yields the optimal low-rank distance-preserving
embedding on which we build the ordering, rather than an ad-hoc proxy,
and a few leading directions suffice because the $\sigma_t^{2}$ weights
concentrate the row geometry in the top of the spectrum.

\subsubsection{Polar spectral ordering}
Although a larger $r$ preserves the row distances more faithfully, in
practice we set $r=2$ and use only the rank-2 truncation\cite{zhang2007svd}, for two
reasons.
\begin{itemize}
  \item \textbf{Sufficiency.} Two directions already capture the
  structure that matters at this stage: the $\sigma_t^{2}$ weights place
  most of the row geometry in the leading modes, and we only need
  a coarse neighborhood order here; the fine intra-wave alignment is handled
  in Section~\ref{sec:stage2}.
  \item \textbf{Linearizability.} The scheduler ultimately needs a
  single one-dimensional execution order, so the coordinates must be
  collapsed onto one axis. Two dimensions form the smallest embedding
  that a polar angle can turn into a total order: every row then has a
  well-defined angle and the rows sort directly, whereas $r\ge3$ admits
  no comparably cheap and rotation-robust linearization and would
  instead require a clustering or space-filling heuristic.
\end{itemize}

As analyzed above, we can represent each row $i$ by a 2-dimensional point $(\sigma_1u_1[i],\sigma_2u_2[i])$, and then sort them by their polar angles. Note that $\sigma_1, \sigma_2 \geq 0$, so they merely reparametrizes the angle monotonically but leaves the resulting order unchanged.  We therefore drop the singular-value
weights and order the rows by the polar angles
\begin{equation}
a_i=\operatorname{atan2}\bigl(u_2[i],u_1[i]\bigr).
\end{equation}
As shown in Figure~\ref{fig:Stage1}, each row $i$ is placed at the point $(u_1[i], u_2[i])$, and the polar angle  unrolls the ring into a one-dimensional sequence where angularly adjacent rows tend to share more K/V blocks.

\subsubsection{Matrix-Free Centered Subspace Iteration}
To obtain $u_1,u_2$, a direct approach would first form
$\widetilde{M}_h$ and apply an SVD, or equivalently an eigendecomposition
of the Gram matrix $\widetilde{M}_h\widetilde{M}_h^{\top}$; both are
computationally expensive. Instead, we compute $u_1,u_2$ with a subspace
power iteration, detailed in Algorithm~\ref{alg:mfcsi}. Crucially, it
never materializes $\widetilde{M}_h$ or the  $M_B\times M_B$ Gram matrix: centering
enters only as a rank-1 correction to two sparse matrix--vector
products,
$\widetilde{M}_h^{\top}X=M_h^{\top}X-\tfrac{1}{M_B}c(\mathbf{1}^{\top}X)$
and $\widetilde{M}_hY=M_hY-\tfrac{1}{M_B}\mathbf{1}(c^{\top}Y)$ with
$c=M_h^{\top}\mathbf{1}$. Each iteration only costs $O(\mathrm{nnz})$. It converges geometrically to the top-2
invariant subspace, and a small fixed $T$ ($5$--$15$) suffices because
the subsequent angular sort is invariant to in-plane rotations and sign
flips of $(u_1,u_2)$.

It is worth emphasizing that the centering step is necessary rather than cosmetic: the embedding coordinates
are measured from the origin, so on the raw mask the leading singular
vector is spent on the row centroid, which tracks the per-row workload
($\sigma_1 u_1[i]=\langle M_h[i],v_1\rangle\approx n_i$) rather than the
K/V-access shape. Centering sends this constant (workload) direction to
a zero singular value ($\mathbf{1}^{\top}\widetilde{M}_h=\mathbf{0}$),
so both retained coordinates $u_1,u_2$ of $\widetilde{M}_h$ carry purely structural information.

\begin{algorithm}[t]
    \caption{Matrix-Free Centered Subspace Iteration (MFCSI)}
    \label{alg:mfcsi}
    \footnotesize
    \begin{algorithmic}[1]
      \REQUIRE binary mask
      $M\in\{0,1\}^{M_B\times N_B}$,
      iterations $T$
      \ENSURE top-$2$ left singular vectors
      $u_1,u_2$ of
      $\widetilde{M}=\left(I-\frac{1}{M_B}\mathbf{1}\mathbf{1}^{\top}\right)M$

      \STATE $c \leftarrow M^{\top}\mathbf{1}$
      \STATE $m \leftarrow M_B$
      \STATE $X \leftarrow$ random orthonormal block of shape $M_B\times 2$

      \STATE \textbf{Subspace power iteration}
      \FOR{$t=1$ to $T$}
        \STATE $Y \leftarrow M^{\top}X-\frac{1}{m}\,c\,(\mathbf{1}^{\top}X)$
        \STATE $Z \leftarrow M Y-\frac{1}{m}\,\mathbf{1}\,(c^{\top}Y)$
        \STATE $X \leftarrow Q$ from the QR factorization $Z=QR$
      \ENDFOR

      \STATE \textbf{Optional: order by singular value}
      \STATE $Y \leftarrow M^{\top}X-\frac{1}{m}\,c\,(\mathbf{1}^{\top}X)$;\quad
      $Z \leftarrow M Y-\frac{1}{m}\,\mathbf{1}\,(c^{\top}Y)$
      \STATE $B \leftarrow X^{\top}Z$
      \STATE $(\Lambda,R)\leftarrow \mathrm{eig}(B)$
      \STATE $X \leftarrow X R$
      \STATE $u_1,u_2 \leftarrow$ columns of $X$

      \RETURN $u_1,u_2$
    \end{algorithmic}
  \end{algorithm}

\begin{figure}
    \centering
    \includegraphics[width=1\linewidth]{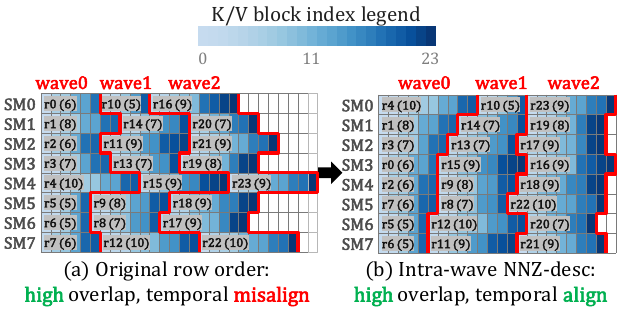}
    \caption{Effect of Stage 2 intra-wave ordering, $r_i(n)$ denotes row $i$ with $n$ nonzero blocks.}
    \label{fig:Stage2}
\end{figure}

\subsection{NNZ-Guided Intra-Wave Temporal Alignment} \label{sec:stage2}
\label{subsec:nnz-reordering}

The SVD-based ordering mentioned above (Stage~1) addresses spatially dispersed K/V accesses by grouping rows with similar access patterns.
However, as discussed in Section~\ref{subsec:two-root-cause}, it does not account for the varying numbers of nonzero blocks across rows.
Figure~\ref{fig:Stage2}(a) illustrates 24 rows scheduled on 8 SMs over 3 waves.
SM0--SM7 initially execute rows $r_0$--$r_7$, but rows with fewer NNZ blocks finish earlier and immediately fetch subsequent rows from the Wave 1.
For example, after completing $r_5$ and $r_6$, the corresponding SMs proceed to $r_9$ and $r_8$, respectively.
This progressive execution skew creates temporal misalignment across waves.
Consequently, although Stage~1 places rows with overlapping K/V accesses close together, their accesses to the same blocks may occur far apart in time, weakening L2 cache reuse.

To mitigate this temporal misalignment, we first sort the rows within each Stage~1 Wave in descending order of NNZ blocks and then schedule them onto the SMs.
As shown in Figure~\ref{fig:Stage2}(b), the first Wave assigns the longest row $r_4$ to SM0 and the shortest row $r_6$ to SM7.
After the shorter rows $r_5$ and $r_6$ complete, their SMs fetch the longest rows in the next Wave, $r_{12}$ and $r_{11}$, respectively.
This long-short complementary assignment at successive Wave boundaries limits the accumulation of execution skew and keeps the SMs approximately aligned.
Because rows in adjacent waves have similar K/V access patterns after Stage~1, temporal alignment allows their shared K/V blocks to be accessed within shorter reuse distances, thereby improving L2 cache locality.

\begin{algorithm}[t]
  \caption{Two-stage reordering}
  \label{alg:WaveAlign-center}
  \footnotesize
  \begin{algorithmic}[1]
    \REQUIRE binary mask
    $M\in\{0,1\}^{H\times M_B\times N_B}$,
    wave size $W$
    \ENSURE per-head row permutations $P_h$

    \FOR{$h=1$ to $H$}

      \STATE \textbf{Stage 1: SVD-based global spatial ordering}
      \STATE Compute the top-$2$ left singular vectors
      $u_1,u_2$  by MFCSI
      \STATE $a_i
      \leftarrow
      \mathrm{atan2}(u_2[i],u_1[i])$
      \STATE $O \leftarrow$ rows sorted by $a_i$
      \STATE Split $O$ into consecutive buckets
      $B_1,\ldots,B_K$ of at most $W$ rows

      \STATE \textbf{Stage 2: NNZ-based intra-wave alignment}
      \STATE $NNZ_i \leftarrow \sum_j M_h[i,j]$
      for all rows $i$
      \FOR{each bucket $B_k$}
        \STATE sort rows in $B_k$ by descending $NNZ_i$
      \ENDFOR

      \STATE $P_h
      \leftarrow$ concatenation of the refined buckets
      $B_1,\ldots,B_K$

    \ENDFOR
  \end{algorithmic}
\end{algorithm}

\subsection{Runtime Coordination}
\label{subsec:runtime-coordination}

\para{Two-stage coordination.}
Sections~\ref{subsec:svd-reordering} and~\ref{subsec:nnz-reordering} improve L2 locality from complementary spatial and temporal dimensions, respectively.
Algorithm~\ref{alg:WaveAlign-center} integrates the two stages.
Given the sparse mask $M$ and wave size $W$, \sysname reorders each attention head independently (Line~1).
It first applies Stage~1 to derive a global row order based on K/V-access similarity (Lines~2--6), and partitions the ordered rows into wave-sized buckets (Line~7).
Within each bucket, Stage~2 sorts rows by descending NNZ count to reduce cross-wave execution skew (Lines~8--12).
Finally, the refined buckets are concatenated to produce the row permutation that can be directly scheduled by the backend kernel (Line~13).

\para{Adaptive reordering skip.}
Row reordering is not always profitable, because its overhead must be amortized by the resulting kernel-time reduction.
We skip reordering in three cases.
First, for short sequences, the entire K/V working set fits in the L2 cache, making row order largely irrelevant to cache reuse.
Second, when the mask is too sparse, the remaining attention workload is insufficient to amortize the reordering cost.
Third, when adjacent rows already exhibit high K/V-access similarity, the original order provides adequate L2 locality.
To make this decision automatically, our adaptive skip module extracts lightweight statistics from the sparse mask and combines them with offline profiles of kernel latency and reordering cost.
Reordering is enabled only when the estimated latency reduction exceeds its overhead.

\section{IMPLEMENTATION}
\label{sec:implementation}

We integrate \sysname into vLLM-Omni~\cite{vllm_omni} as a lightweight plug-in between sparse-mask construction and backend-kernel execution.

\para{Reordering operator.}
We introduce two additional operators.
The first implements the two-stage row reordering described in Algorithm~\ref{alg:WaveAlign-center}, applying the resulting permutation to the query blocks and sparse mask before attention kernel.
The second applies the inverse permutation to restore the attention outputs to their original row order.
Both operators are composed from PyTorch tensor primitives and registered in vLLM-Omni, allowing them to be invoked transparently throughout the inference pipeline.

\para{FlashInfer per-head scheduling.}
On the A40 platform, we use FlashInfer~\cite{flashinfer} as the backend kernel. Its native scheduler prioritizes the K/V-head dimension, allowing Q-block computations from different heads to run concurrently across SMs.
Because each head accesses a separate set of K/V blocks without cross-head overlap, concurrent execution only increases contention in the shared L2 cache.
Following FlashAttention4~\cite{flashattention4}, we instead prioritize the query-sequence dimension and execute K/V heads sequentially.
This exposes only one head's K/V working set at a time, reducing L2 contention and improving intra-head locality.
Section~\ref{subsec:evaluation-e2e} and ~\ref{subsec:evaluation-overhead} evaluate the effectiveness of this scheduling modification.

\begin{figure*}[!t]
  \centering
  \includegraphics[width=\textwidth]{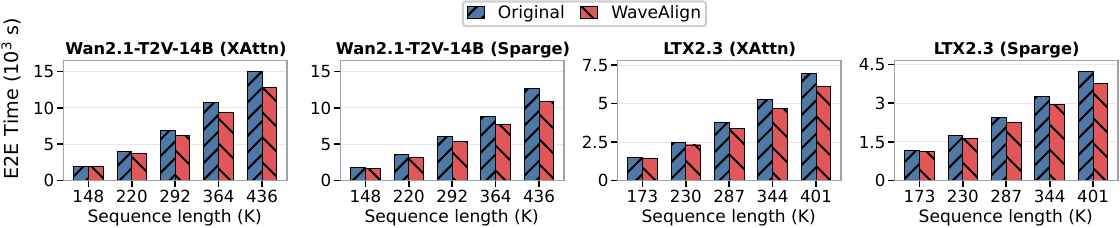}
  \caption{End-to-end video generation latency with and without \sysname on Wan2.1-T2V-14B and LTX2.3 across various sequence lengths atop H100.}
  \label{fig:e2e-wan-ltx-speedup}
\end{figure*}

\section{Evaluation}

\subsection{Experimental Setup}
\label{subsec:evaluation-setup}

\noindent\textbf{Hardware and software platforms.} 
We evaluate \sysname on two NVIDIA GPU platforms.  
The H100 GPU provides 80GB of HBM3 memory and 50MB L2 cache shared by 132 SMs.
The A40 GPU provides 48GB of GDDR6 memory and 6MB L2 cache shared by 84 SMs.
The primary software stack includes CUDA 13.0~\cite{cuda130}, PyTorch 2.11~\cite{pytorch211}, NsightCompute 2025.3.1~\cite{nsight_compute_2025_3_1}, vLLM-Omni 0.20.0~\cite{vllm_omni}.
We use FlashAttention4~\cite{flashattention4} as the backend kernel on H100 GPU and FlashInfer~\cite{flashinfer} on A40 GPU, as FlashAttention4 depends on the Hopper and Blackwell architecture and does not support Ampere GPUs.

\noindent\textbf{Models.} We evaluate \sysname on two representative open-source Text-to-Video (T2V) model families, Wan2.1~\cite{wan21_repo} and LTX2.3~\cite{ltxvideo}. Our H100 experiments use Wan2.1-T2V-14B~\cite{wan21_t2v_14b_modelcard} and LTX2.3-22B, whereas A40 uses the smaller Wan2.1-T2V-1.3B~\cite{wan21_t2v_1p3b_modelcard} due to the limited memory capacity.

\noindent\textbf{Workloads.} We use prompts from the Penguin Video Benchmark~\cite{penguin_video_benchmark} to evaluate video generation. 
To construct workloads with different sequence lengths, we vary the number of frames while keeping the resolution fixed.
For Wan2.1-T2V-14B, we use a resolution of $720{\times}1280$ and vary the frames from 161 to 481, yielding sequences of 148K--436K tokens. 
For LTX2.3-22B, the resolution is $1216{\times}1920$, and we vary the range of frames from 601 to 1401, yielding sequences of 173K--401K. 
For Wan2.1-T2V-1.3B, we set resolution as $720{\times}1280$ and frames from 121 to 241, yielding sequences of 112K--220K.

\noindent\textbf{Baselines and configurations.} 
Our baselines are four sparse-attention methods without applying \sysname under default sparse configurations: 
XAttention~\cite{xattention} with threshold~0.9, SpargeAttn~\cite{spargeattention} and MInference~\cite{minference} with pre-autotuned config and FlexPrefill~\cite{flexprefill} with $g=0.60$. 
We use these configurations for the kernel-level evaluation of four methods. For the end-to-end evaluation, however, we exclude MInference and FlexPrefill because their overhead of mask construction is prohibitively high in our T2V workloads.
For FlashAttention4 on H100, we use a block size of $128{\times}128$ tokens and one resident cooperative thread array (CTA) per SM across 132 SMs, yielding a wave size of 132. 
For FlashInfer on A40, we use a block size of $64{\times}64$ tokens and two resident CTAs per SM across 84 SMs, yielding a wave size of 168.

\noindent\textbf{Metrics.} We report end-to-end generation latency and sparse-attention kernel latency, together with speedup relative to the corresponding baselines. 
We separately measure the total reordering overhead, including permutation construction, query and mask gathering, and inverse for output. 
To explain the latency reduction, we report L2 cache hit ratio and HBM read traffic. 
Finally, to verify that \sysname preserves the sparse-attention semantics and generated video quality, we report PSNR, SSIM, LPIPS, and the VBench~\cite{huang_vbench} metrics Imaging Quality and Subject Consistency.

\subsection{End-to-End Video Generation Latency}
\label{subsec:evaluation-e2e}

\begin{figure}[t]
  \centering
  \includegraphics[width=\columnwidth]{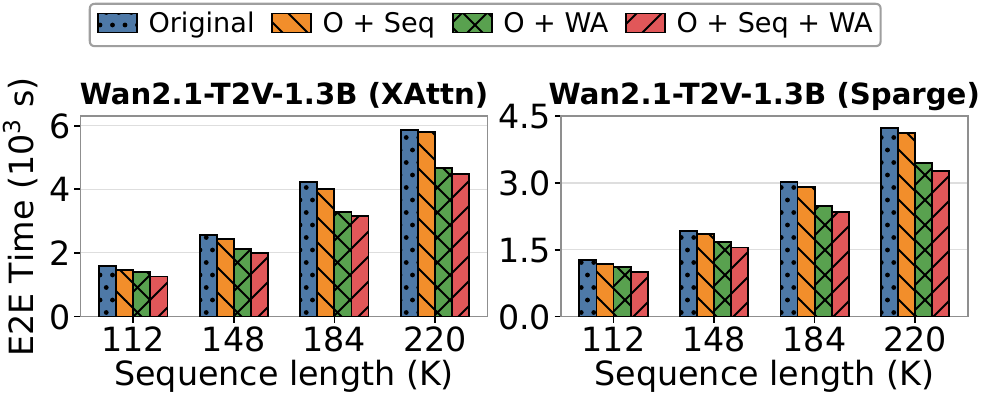}
  \caption{E2E video generation latency of Wan2.1-T2V-1.3B on A40/FlashInfer under different combinations of head scheduling and \sysname.}
  \label{fig:a40-flashinfer-variants}
\end{figure}

\noindent\textbf{H100 results.}
Figure~\ref{fig:e2e-wan-ltx-speedup} reports the end-to-end video generation latency of XAttention and SpargeAttn with and without \sysname on Wan2.1-T2V-14B and LTX2.3-22B.
In the original baselines, sparse attention dominates end-to-end latency, accounting for 65.8\%--88.0\% on Wan2.1-T2V-14B and 58.1\%--84.4\% on LTX2.3-22B. Consequently, \sysname's kernel-level improvements translate into consistent end-to-end speedups across both models, both sparse-attention methods, and all evaluated sequence lengths, reaching up to $1.17\times$ on Wan2.1-T2V-14B and $1.14\times$ on LTX2.3-22B.
The gains generally increase with sequence length because the growing K/V working set increasingly exceeds the fixed L2 cache capacity, leading to more severe cache thrashing and heavy HBM accesses under the original row order.

\noindent\textbf{A40 results.}
Figure~\ref{fig:a40-flashinfer-variants} reports the end-to-end generation latency of XAttention and Sparge on Wan2.1-T2V-1.3B under four configurations: FlashInfer's native concurrent-head execution (\textit{Original}), the sequential-head scheduler described in Section~\ref{sec:implementation} (\textit{O+Seq}), \sysname applied to the native scheduler (\textit{O+WA}), and the combination of both optimizations (\textit{O+Seq+WA}).
Compared with \textit{Original}, \textit{O+Seq} alone achieves a speedup of up to 1.09$\times$, demonstrating the benefit of our sequential-head implementation. 
Applying \sysname directly to \textit{Original} yields a speedup of up to 1.29$\times$. Combining the two optimizations further improves upon \textit{O+WA} by up to 11.0\%, with \textit{O+Seq+WA} consistently achieving the lowest latency.
The two optimizations improve cache behavior in complementary ways: sequential-head execution prevents different heads from simultaneously competing for the A40's 6MB L2 cache, while \sysname further increases the L2 hit ratio by improving intra-head K/V reuse.

\begin{figure*}
    \centering
    \includegraphics[width=1\textwidth]{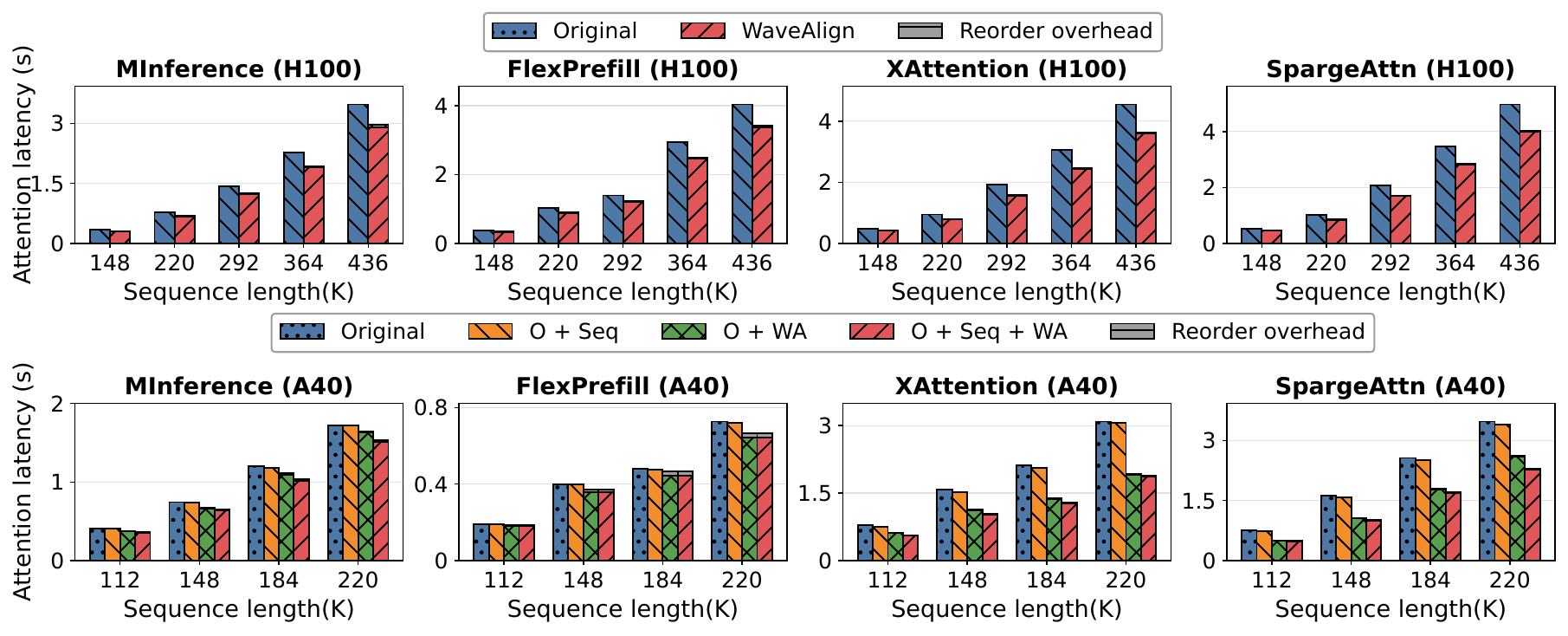}
    \caption{Sparse-attention kernel latency on H100 (top) and A40 (bottom). For \sysname configurations, the complete reordering overhead is shown separately above the kernel latency.}
    \label{fig:overhead-breakdown}
\end{figure*}

At comparable sequence lengths, the gains on A40 are larger than those on H100 because the A40's much smaller L2 cache becomes capacity-constrained earlier, making cache locality important even for shorter sequences.

\subsection{Kernel Latency Breakdown}
\label{subsec:evaluation-overhead}

\sysname optimizes only sparse attention, so its end-to-end speedup comes entirely from reduced attention latency. We next isolate the attention computation and analyze its kernel-level latency on H100 and A40.

\noindent\textbf{H100 results.}
Figure~\ref{fig:overhead-breakdown} illustrates the H100 sparse-attention kernel latency of MInference, FlexPrefill, XAttention, and SpargeAttn using real QKV tensors captured from Wan2.1-T2V-14B. 
Each \sysname bar separates the optimized kernel latency from the complete reordering overhead. 
Including this overhead, \sysname consistently outperforms the original kernel across all four methods and sequence lengths, achieving speedups of $1.092\times$--$1.181\times$ for MInference, $1.088\times$--$1.179\times$ for FlexPrefill, $1.124\times$--$1.252\times$ for XAttention, and $1.125\times$--$1.234\times$ for SpargeAttn. 
Reordering accounts for only 1.1\%--3.2\% of the total \sysname attention latency and is consistently outweighed by the kernel-time reduction. 

The gains increase with sequence length as the growing K/V working set exceeds L2 capacity, amplifying cache evictions and repeated HBM accesses that \sysname mitigates through shorter reuse distances. 
The gains are also larger for XAttention and SpargeAttn because their fully content-dependent masks produce more irregular K/V accesses than the structured masks of MInference and FlexPrefill.

\noindent\textbf{A40 results.}
The A40 panels in Figure~\ref{fig:overhead-breakdown} report the kernel latency of the same four methods using real QKV tensors captured from Wan2.1-T2V-1.3B. 
We evaluate parallel- and sequential-head execution, each with and without \sysname, using FlashInfer's native parallel-head scheduler as the baseline. 
The legend follows the naming convention in Section~\ref{subsec:evaluation-e2e}, and all reported \sysname speedups include the complete reordering overhead.
Across the four sequence lengths, \textit{O+WA} achieves maximum speedups over \textit{Original} of $1.104\times$, $1.090\times$, $1.599\times$, and $1.520\times$ for MInference, FlexPrefill, XAttention, and SpargeAttn, respectively. 
Relative to \textit{O+Seq}, \textit{O+Seq+WA} achieves maximum speedups of $1.133\times$, $1.077\times$, $1.614\times$, and $1.556\times$, respectively. 
Compared with \textit{Original}, \textit{O+Seq+WA} achieves maximum speedups of $1.158\times$, $1.090\times$, $1.635\times$, and $1.601\times$, and consistently provides the lowest total attention latency.
\sysname improves both parallel- and sequential-head execution across all four methods, with its 7.5--29.7ms reordering overhead consistently outweighed by the kernel-time reduction. Sequential-head scheduling reduces cross-head L2 contention, while \sysname improves intra-head K/V locality, making \textit{O+Seq+WA} the most effective configuration.

Compared with H100, A40 is more sensitive to row ordering. While \sysname achieves up to $1.252\times$ speedup on H100, \textit{O+WA} reaches $1.599\times$ and $1.520\times$ for XAttention and SpargeAttn on A40, respectively, even at shorter sequence lengths. This is because the A40's smaller L2 cache (6MB versus 50MB on H100) becomes capacity-constrained earlier, amplifying the benefits of row reordering for irregular, content-dependent masks.

\begin{figure}[t]
  \centering
  \includegraphics[width=1\linewidth]{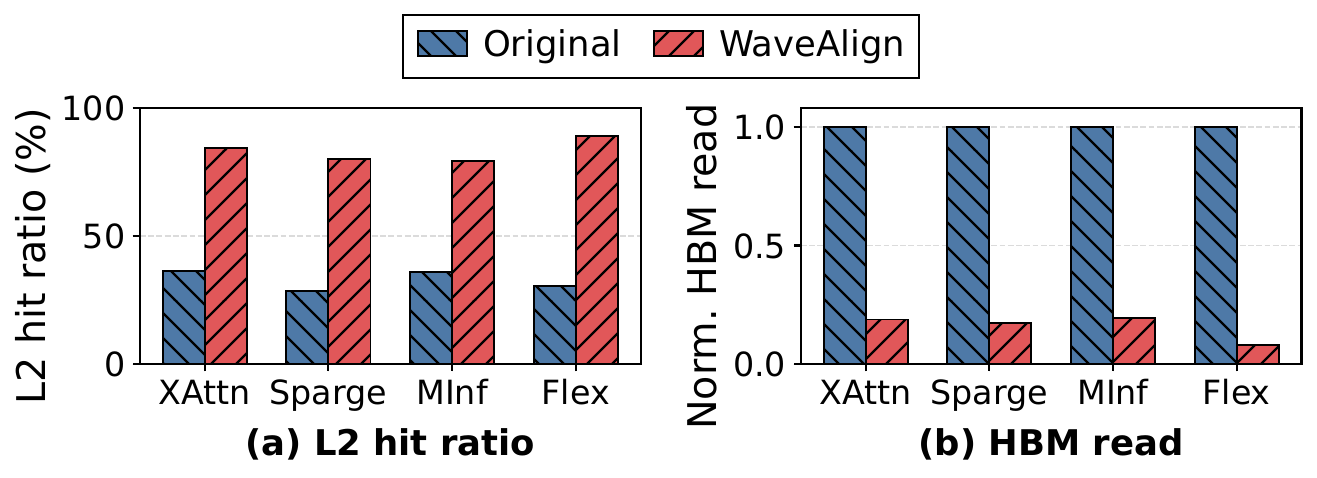}
  \caption{L2 cache hit ratio and HBM read traffic on H100/FlashAttention-4 with and without \sysname across four sparse-attention methods. HBM reads are normalized to the corresponding no-reorder baseline, and all results report the median of three NCU runs.}
  \label{fig:evaluation-kernel-locality}
\end{figure}

\subsection{L2 Cache Locality and HBM Traffic}
\label{subsec:evaluation-kernel}
To identify the source of the kernel speedup, we use NVIDIA Nsight Compute (NCU)~\cite{nsight_compute_2025_3_1} to profile FlashAttention4 on H100 using a Wan2.1-T2V-14B workload with 401 frames at $720{\times}1280$, corresponding to 364K visual tokens.
Figure~\ref{fig:evaluation-kernel-locality} shows consistent locality improvements across all four sparse methods: the L2 cache hit ratio increases from 28.48\%--36.35\% to 79.38\%--89.06\%, while HBM reads decrease by 80.71\%--92.11\%.
In particular, XAttention and SpargeAttn have the lowest baseline hit ratios because their fully content-dependent masks produce more dynamic and irregular K/V access patterns. 
\sysname raises their hit ratios from 36.35\% and 28.48\% to 84.17\% and 80.04\%, respectively.

\begin{figure*}[t]
  \centering
  \includegraphics[width=1\textwidth]{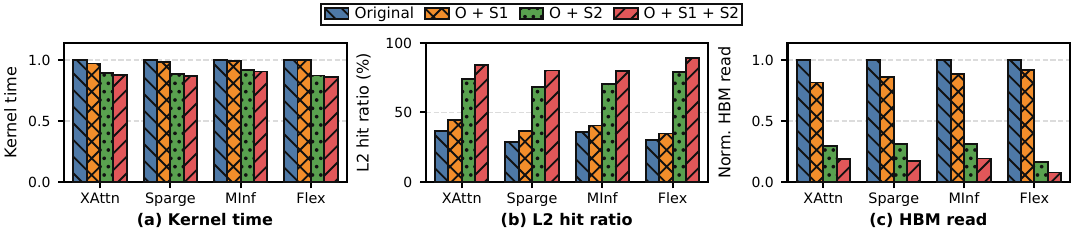}
  \caption{Ablation of \sysname's two-stage reorder across four sparse-attention methods under four cumulative configurations: Original, O+S1, O+S2 and O+S1+S2. S1 and S2 denote global spatial ordering and intra-wave temporal alignment, respectively. Kernel time and HBM reads are normalized to Original.}
  \label{fig:abalation}
\end{figure*}

Since \sysname preserves both the sparse mask and computation, these results show that its speedup comes from improved K/V reuse: shorter reuse distances keep more K/V blocks in L2 and avoid repeated HBM accesses.

\subsection{Ablation Study}
\label{subsec:evaluation-ablation}

Figure~\ref{fig:abalation} evaluates the two stages of \sysname on H100 with FlashAttention4 using a Wan2.1-T2V-14B workload of 401 frames at $720{\times}1280$, corresponding to 364K visual tokens. 
We evaluate XAttention, SpargeAttn, MInference, and FlexPrefill under four cumulative configurations: \textit{Original}, \textit{O+S1}, \textit{O+S2}, and \textit{O+S1+S2}. 
Stage~1 applies the global spatial row ordering described in Section~\ref{subsec:svd-reordering}, while Stage~2 adds the intra-wave NNZ-based refinement described in Section~\ref{sec:stage2}. We use an execution-wave size of 132 rows, matching the number of SMs on H100.

Stage~1 alone provides modest improvements, achieving up to $1.0319\times$ kernel speedup, increasing the L2 hit ratio from 28.48\%--36.35\% to 34.88\%--44.47\%, and reducing HBM read traffic by 8.71\%--18.61\%. Stage~2 alone delivers substantially larger gains. Across the four methods, \textit{O+S2} achieves $1.091\times$--$1.147\times$ kernel speedup, increases the L2 hit ratio to 68.35\%--78.71\%, and reduces HBM read traffic by 68.89\%--83.67\% relative to \textit{Original}.
Combining both stages consistently produces the best results. Relative to \textit{Original}, \textit{O+S1+S2} accelerates the kernel by up to $1.17\times$, raises the L2 hit ratio to 79.38\%--89.06\%, and reduces HBM traffic by up to 92.11\%. 

The two stages improve L2 reuse along complementary dimensions. Stage~1 groups rows with similar K/V access patterns into nearby waves, increasing the likelihood that different SMs request the same blocks. 
However, uneven row workloads can separate these requests in time and prevent cache reuse. Stage~2 sorts rows by descending NNZ to align CTA progress and shorten reuse distances, keeping shared K/V blocks in L2 for subsequent accesses. 
Stage~1 establishes spatial overlap, while Stage~2 converts it into temporal reuse; both are therefore necessary for effective L2 reuse.

\subsection{Other Factors}

\begin{figure}[t]
  \centering
  \includegraphics[width=\columnwidth]{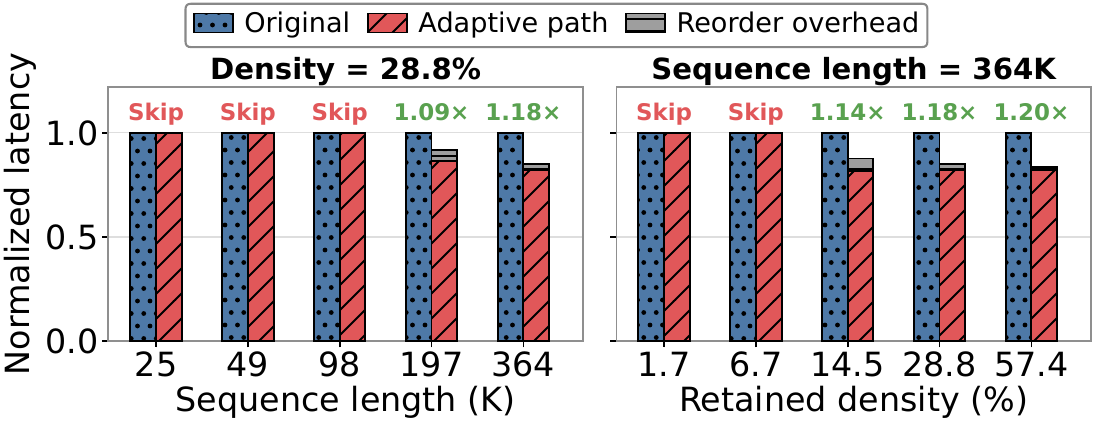}
  \caption{Adaptive reordering skip for XAttention with FlashAttention4 on Wan2.1-T2V-14B under different sequence lengths and retained densities.}
  \label{fig:adaptive_skip}
\end{figure}

\para{Adaptive reordering skip.}
Figure~\ref{fig:adaptive_skip} demonstrates the necessity of adaptive reordering proposed in Section~\ref{subsec:runtime-coordination} by reporting kernel-level latency, including the reordering overhead, for XAttention on Wan2.1-T2V-14B with FlashAttention4.
At short sequence lengths of 25K, 49K, and 98K, the kernel-time reduction is insufficient to amortize the reordering cost, so \sysname skips reordering and falls back to the default sparse-attention path, corresponding to a $1.0\times$ speedup over the baseline.
From 197K onward, the growing K/V working set makes cache locality increasingly important, and reordering becomes profitable, delivering up to $1.18\times$ speedup.
Retained density exhibits a similar trend: \sysname skips reordering at 1.7\% and 6.7\% density, but enables it from 14.5\% onward, achieving $1.14\times$--$1.20\times$ speedup. These results show that adaptive skipping prevents reordering overhead from causing regressions while preserving its benefits in profitable regimes.

\begin{figure}[t]
  \centering
  \includegraphics[width=0.97\columnwidth]{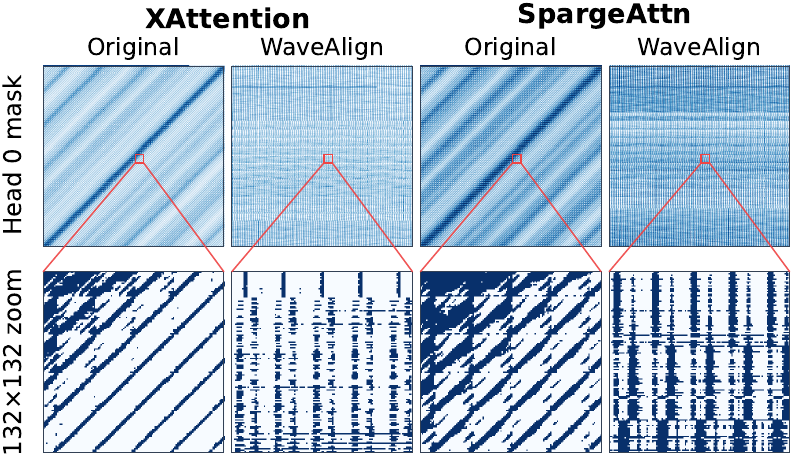}
  \caption{Head-0 sparse masks of XAttention and SpargeAttn before and after \sysname. The top row shows the full masks, with local regions marked in red; the bottom shows the corresponding $132{\times}132$ zoomed windows.}
  \label{fig:mask-reorder-shape-gallery}
\end{figure}

\para{Visualization of sparse mask.} 
Figure~\ref{fig:mask-reorder-shape-gallery} visualizes the head-0 sparse masks of XAttention and SpargeAttn on Wan2.1-T2V-14B before and after \sysname reordering. Each zoomed region spans 132 query rows, corresponding to one execution wave on H100, and 132 K/V-block columns. 
Before reordering, the masks are dominated by diagonal patterns, and rows within the same wave exhibit little column-wise overlap, indicating that different SMs rarely access the same K/V blocks. 
After reordering, more vertically aligned nonzero blocks appear, showing that rows executed within the same wave are more likely to share K/V accesses. 
Since \sysname preserves each row's selected blocks and changes only the execution order, this increased column-wise overlap directly improves cross-SM L2 reuse.

\para{Compare with other reorder algorithms.}
Table~\ref{tab:xattn-reorder-baselines} compares \sysname with four representative locality-aware sparse matrix multiplication reordering methods reviewed in Section~\ref{sec:related-work}, using the configuration in Section~\ref{subsec:evaluation-kernel}. These methods expose a trade-off between locality and overhead. Clustering has the lowest overhead among these baselines but barely raises the L2 hit ratio from 17.57\% to 19.29\%. Hypergraph, MinHash, and LSH substantially improve L2 locality, but their preprocessing costs outweigh the kernel savings, resulting in inference slowdowns. 
In contrast, \sysname raises the L2 hit ratio to 76.72\% with only 19.85ms of overhead, retaining a $1.210\times$ net speedup. It is therefore the only evaluated method that combines effective locality improvement with practical online overhead.

\begin{table}[t]
  \centering
  \scriptsize
  \setlength{\tabcolsep}{1.3pt}
  \renewcommand{\arraystretch}{1.08}
  \caption{Comparison with other reordering algorithms.}
  \label{tab:xattn-reorder-baselines}
  \begin{tabular*}{\columnwidth}{@{\extracolsep{\fill}}lrrrrrr@{}}
    \toprule
    Method &
    \shortstack{Time\\(ms)} &
    \shortstack{Speedup\\w/o overhead} &
    \shortstack{Speedup\\w/ overhead} &
    \shortstack{Overhead\\(ms)} &
    \shortstack{HBM SOL\\(\%)} &
    \shortstack{L2 Hit\\(\%)} \\
    \midrule
    Original  & 2723.99 & 1.000 & 1.000 &      0.00 & 83.98 & 17.57 \\
    \sysname    & 2234.37 & 1.221 & 1.210 &     19.85 & 15.37 & 76.72 \\
    Hypergraph & 2198.40 & 1.236 & 0.005 & 535435.34 & 10.38 & 83.36 \\
    MinHash     & 2257.93 & 1.208 & 0.715 &   1556.40 & 19.24 & 72.26 \\
    LSH         & 2259.70 & 1.203 & 0.715 &   1540.98 & 20.90 & 70.30 \\
    Clustering  & 2707.63 & 1.008 & 0.961 &   133.07 & 81.65 & 19.29 \\ 
    \bottomrule
  \end{tabular*}
\end{table}

\begin{figure}[t]
  \centering
  \includegraphics[width=0.97\columnwidth]{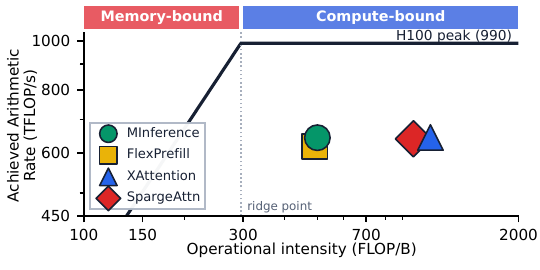}
  \caption{Roofline analysis on attention kernels with \sysname of Wan2.1-14B model with 364K tokens on H100.}
  \label{fig:wavealign-roofline}
\end{figure}
\para{Roofline analysis.}  
Compared with Figure~\ref{fig:roofline471K}, Figure~\ref{fig:wavealign-roofline} shows that \sysname increases operational intensity and achieved arithmetic rate, moving all four sparse-attention kernels across the ridge point from memory-bandwidth-bound to compute-bound execution.

\begin{table}[t]
  \centering
  \scriptsize
  \setlength{\tabcolsep}{4pt}
  \renewcommand{\arraystretch}{1.05}
  \caption{Quality and consistency metrics before and after \sysname.}
  \label{tab:quality-consistency}
  \resizebox{\columnwidth}{!}{%
  \begin{tabular}{@{}llrrrrrr@{}}
    \toprule
    Model & Sparse method & Density & PSNR & SSIM & LPIPS & ImgQual & SubCons \\
    \midrule
    Wan2.1-T2V-14B & Dense & 1.000 & -- & -- & -- & 0.695 & 0.963 \\
                    & XAttn & 0.507 & 17.498 & 0.581 & 0.284 & 0.686 & 0.839 \\
                    & XAttn+\sysname & 0.507 & 17.498 & 0.581 & 0.284 & 0.686 & 0.839 \\
                    & Sparge & 0.559 & 19.161 & 0.675 & 0.224 & 0.694 & 0.956 \\
                    & Sparge+\sysname & 0.559 & 19.161 & 0.675 & 0.224 & 0.694 & 0.956 \\
    \addlinespace[1pt]
    \midrule
    LTX2.3-22B & Dense & 1.000 & -- & -- & -- & 0.581 & 0.912 \\
           & XAttn & 0.599 & 27.683 & 0.910 & 0.094 & 0.594 & 0.899 \\
           & XAttn+\sysname & 0.599 & 27.683 & 0.910 & 0.094 & 0.594 & 0.899 \\
           & Sparge & 0.588 & 26.485 & 0.885 & 0.115 & 0.589 & 0.898 \\
           & Sparge+\sysname & 0.588 & 26.485 & 0.885 & 0.115 & 0.589 & 0.898 \\
    \bottomrule
  \end{tabular}%
  }
\end{table}

\para{Video generation accuracy.}
Table~\ref{tab:quality-consistency} evaluates video generation quality on the Penguin Video Benchmark~\cite{penguin_video_benchmark} using Wan2.1-T2V-14B and LTX2.3-22B with XAttention and SpargeAttn. All videos are generated at 480p with 81 frames. Across both models and sparse attention methods, the original and \sysname variants achieve identical results on all reported metrics. This is because \sysname only reorders the mask rows and their corresponding queries, while preserving the selected K/V blocks and the backend kernel's per-row computation. The outputs are then restored to their original order, leaving the sparse attention semantics and inference accuracy unchanged.

\section{Related Work}
\label{sec:related-work}

\para{Sparse attention for video generation.}
This topic has attracted growing interest. \textit{Fixed-pattern methods} reuse predefined layouts across inputs and denoising steps. Rule-based approaches, including LongNet~\cite{longnet}, LogSparse~\cite{logsparse}, STA~\cite{sta}, Radial Attention~\cite{radial_attention}, and LVSA~\cite{lvsa}, exploit spatiotemporal locality, whereas profile-guided methods, such as SVG~\cite{svg} and Sparse-vDiT~\cite{sparse_vdit}, select head- or layer-specific patterns from representative attention profiles. \textit{Content-adaptive methods}, such as AdaSpa~\cite{adaspa} and SVG2~\cite{sparsevideogen2}, identify important blocks or tokens from the current input to balance sparsity and generation quality. Complementary to these methods, \sysname reorders rows after mask construction to improve kernel efficiency without changing the selected blocks.

\para{GPU kernels for sparse attention.}
To accelerate attention, prior kernel optimizations fall into two groups. The FlashAttention series~\cite{flashattention,flashattention2,flashattention4} improve kernel efficiency through IO-aware tiling, online softmax, and GPU pipelining. 
Sparse or flexible attention kernels, including FlashInfer~\cite{flashinfer}, FlexAttention~\cite{flexattention}, FlashMask~\cite{flashmask}, Block Sparse FlashAttention~\cite{block_sparse_flashattention}, and FSA~\cite{fsa}, further support diverse sparse masks. 
\sysname complements these backends by reordering the mask to unlock their full performance.

\para{Locality-aware sparse matrix reordering.}
Reordering is widely used to improve locality in sparse matrix multiplication, with representative approaches based on LSH-based row reordering~\cite{jiang_lsh_spmm}, hypergraph partitioning~\cite{hypergraph_spgemm}, MinHash-based ordering~\cite{chierichetti_shingle}, and clustering~\cite{bootes}. 
However, our prior evaluation shows that the high preprocessing overhead makes them unsuitable for runtime-generated sparse attention masks. 
To the best of our knowledge, \sysname provides a fast and effective reordering method tailored to sparse attention.

\section{CONCLUSION}

\noindent \sysname is an efficient reordering framework for accelerating dynamic sparse attention in long-video generation. 
By jointly improving spatial KV locality and temporal workload alignment, it increases L2 cache reuse and reduces HBM traffic without modifying the sparse attention algorithm. 
\sysname reduces end-to-end video generation latency by up to 1.17× and sparse attention latency by up to 1.25× over strong baselines across multiple models and algorithms.

\bibliographystyle{IEEEtranS}
\bibliography{refs}
\appendices

\section*{AI Use}

ChatGPT-5.6 Extra High was used solely to polish the manuscript's language and improve readability. All other aspects of the work, including problem identification, algorithm and system design, implementation, experimental evaluation, and analysis of results, were conducted independently by the authors without AI assistance.
\end{document}